# Single crystal growth and physical property characterization of PbTaSe$_2$ as a noncentrosymmetric type-II superconductor

**Raman Sankar**[a,b,*] **G. Narsinga Rao**[b], **I. Panneer Muthuselvam**[b], **Guang Bian**[c], **Hao Zheng**[c], **G. Peng-Jen Chen**[a,d,e], **Tay-Rong Chang**[f], **Suyang Xu**[c,], **G. Senthil Murgan**[b], **Chun-Hsuan Lin**[a], **Wei-Li Lee**[a], **Horng-Tay Jeng**[a,f], **M. Zahid Hasan**[c], **and Fang-Cheng Chou**[b,g,h,*]

[a]*Institute of Atomic and molecular Sciences, Academia Sinica, Taipei 10617, Taiwan*
[b]*Center for Condensed Matter Sciences, National Taiwan University, Taipei 10617, Taiwan*
[c]*Laboratory for Topological Quantum Matter and Spectroscopy (B7), Department of Physics, Princeton University, Princeton, New Jersey 08544, USA*
[d]*Department of Physics, National Taiwan University, Taipei 10617, Taiwan*
[f]*Nano Science and Technology Program, Taiwan International Graduate Program,*
[e]*Department of Physics, National Tsing Hua University, Hsinchu 30013, Taiwan*
[g]*National Synchrotron Radiation Research Center, Hsinchu 30076, Taiwan*
[h]*Taiwan Consortium of Emergent Crystalline Materials, Ministry of Science and Technology, Taipei 10622, Taiwan*

*Corresponding Author(s)**

(**1.** *Dr. Raman Sankar*
*Institute of Physics, Academia Sinica,*
*Taipei 10617, Taiwan*
*Email id: sankarndf@gmail.com*

**2.** *Prof. Fangcheng Chou*
*Center for Condensed Matter Sciences,*
*National Taiwan University, Taipei 10617, Taiwan*
*Email id: fcchou@ntu.edu.tw*)

The single crystal growth and superconducting properties of PbTaSe$_2$ with non-centrosymmetric crystal structure is reported. Using the chemical vapor transport (CVT) technique, PbTaSe$_2$ crystallizes in a layered structure and the crystal symmetry has been shown belonging to a non-centrosymmetric space group $P\bar{6}m2$ confirmed by the consistent band picture near the Fermi level between the angle-resolved photoemission spectrum (ARPES) and theoretical calculations. Superconductivity with T$_c$ =3.83±0.02 K has been characterized fully with electrical resistivity ρ(T), magnetic susceptibility χ(T), and specific heat C(T) measurements using single crystal samples. The superconducting anisotropy, electron-phonon coupling λ$_{ep}$, superconducting energy gap Δ$_0$, and the specific heat jump ΔC/γT$_c$ at T$_c$ confirms that PbTaSe$_2$ can be categorized as a weakly coupled type-II superconductor.

**1.** Introduction

The discovery of superconductivity in the non-centrosymmetric (NCS) materials, such as heavy fermion CePt$_3$Si and half-Heusler YPtBi compounds, has initiated widespread research activities in the field of unconventional superconductors.[1-3] The absence of an inversion symmetry creates an asymmetric electric field gradient, which splits the two-fold spin degenerated electronic band into two non-degenerate separated bands.[4,5] In these non-centrosymmetric superconductors (NCS), the new cooper paring states, i.e., coexistence of even and odd parity Cooper pair states are predicted, instead of the conventional type of Cooper pairing in either spin-singlet or spin-triplet.[4] The NCS superconductors have been identified in different materials groups, including heavy fermion, compounds with long range magnetic ordering, such as CePt$_3$Si, CeRhSi$_3$ and UIr,[1-3,6,7] metallic compounds of strong electronic correlation,[9] and weakly correlated materials like Mo$_3$Al$_2$C and Li$_2$(Pd$_{1-x}$Pt$_x$)$_3$B).[10-12]

Recently, Ali *et. al.* reported the existence of superconductivity in a non-centrosymmetric polycrystalline PbTaSe$_2$ sample, which is derived from the typical transition metal dichalcogenide TaSe$_2$ with Pb atoms intercalated between the original van der Waals gap.[13] A comparative study between the single crystal and polycrystalline forms is desirable to learn more about the expected anisotropic physical properties of the layered compound using single crystal samples. In particular, the claim of non-centrosymmetric structure must be confirmed with single crystal. In this article, we report the successful growth of PbTaSe$_2$ single crystal sample using a chemical vapor transport (CVT) technique. Magnetic susceptibility, transport and specific heat measurement results clearly indicate that single crystal PbTaSe$_2$ is superconducting with a superconducting transition temperature of T$_c$ = 3.8 K, and the crystal structure is consistent to a non-centrosymmetric space group of $P\bar{6}m2$ (No.187).

## 2. Result and discussions

Figure 1(a) shows the powder x-ray diffraction (XRD) pattern and Rietveld refinement result of the pulverized as-grown PbTaSe$_2$ single crystals. All diffraction peaks can be indexed with a space group of $P\bar{6}m2$ (No. 187) or $P6/mmm$ (No. 191) equally well statistically.[14] The only difference is that the former has Ta atoms order at the (1/2, 2/3, 1/2) site in $P\bar{6}m2$ without centrosymmetry, and the later has a disordered distribution of Ta atoms of average half occupancy at the (1/3, 2/3, 1/2) site in $P6/mmm$ being centrosymmetric. The structural Rietveld refinement on the XRD pattern gives the lattice constants of $a = 3.4474$ Å and $c = 9.3528$ Å when indexed with $P\bar{6}m2$ space group, which are consistent to the reported values in the literature.[14] The obtained atomic parameters and the site occupancy for space groups $P6/mmm$ and $P\bar{6}m2$ are summarized in Table 1. It is very difficult to distinguish these two space groups based on XRD data alone. The obtained single crystals of size ~ 7 x 5 x 2 mm$^3$ can be cleaved easily to produce flat surfaces, as shown in the inset of Figure 1(a). Figure 1(b) illustrates the XRD pattern of the cleaved surface with diffraction peaks indexed with preferred orientation (00$l$).

While Pb:Ta=1:1 and Ta-Se are coordinated in a trigonal prismatic coordination within each layer, the Pb atoms must sit directly in between Se atoms across layers, which leads to two possible arrangements of Ta atoms. The space group $P6/mmm$ implies that Ta atoms sit at the (1/3, 2/3, 1/2) sites of half occupancy, i.e., Ta is distributed evenly statistically without ordering within each TiSe$_2$ layer (Fig. 2(a) and 2(b)). On the other hand, the space group $P\bar{6}m2$ implies that Ta atoms sit at the (1/3, 2/3, 1/2) sites with full occupancy, i.e., the Ta is ordered within each TiSe$_2$ layer (Fig. 2(c) and 2(d)). Interestingly, the former has a centrosymmetry but the latter lacks of centrosymmetry via an spatial inversion operation of **r** ➔ **-r**. The stacking of the TaSe$_2$ slabs in the pristine 2H-TaSe$_2$ shows the closely packed Se atoms are 180 degrees reversed within each unit, i.e., there are two TaSe$_2$ slabs per unit $c$. On the other hand, PbTaSe$_2$ has one Pb and one TaSe$_2$ trigonal prismatic slab per unit, and the Pb atoms are closely packed in parallel to the Se hexagonal close packing per layer. The stoichiometric chemical composition 1:1:2 for PbTaSe$_2$ single crystal has been confirmed by both the XRD refinement and the chemical analysis of electron probe microanalysis (EPMA) showning Pb:Ta:Se=0.99 : 1 : 2.01 within experimental error. The photoemission core level scan was used to check the chemical composition of the target sample, as shown in Figure 3. Clear Pb-5d, Ta-4f and Se-3d core level peaks were observed in the photoemission spectrum, which confirms the correct chemical composition in our PbTaSe$_2$ single crystal samples and in consistent to the EPMA chemical analysis.

While the XRD lacks the resolution to distinguish whether the PbTaSe$_2$ has a better fitting to the space group of either *P6/mmm* or *P$\bar{6}$m2*, the crystal surface profile has been examined with scanning tunneling microscopy (STM) in detail. PbTaSe2 crystals were prepared for STM measurements by *in vacuo* cleavage at room temperature, in a preparation chamber with a base pressure lower than $1 \times 10^{-10}$ mbar, before transfering to an STM chamber at a pressure lower than $5 \times 10^{-11}$ mbar. STM measurements were performed at a temperature of 4.6 K, using an Omicron LT-STM and an electrochemically etched tungsten tip. The STM image was obtain with a bias voltage of -0.5 V and tunneling current 0.7 nA. Figure 4 shows the constant-current image represents the typical flat surface with highly ordered atomic lattice with the inter-bright spot distance to be near ~3.2 Å. Assuming the bright spots represent the Pb atoms, it is found that the intensity profile represents a crsytal structure in space group (*P$\bar{6}$m2*) better, especially when the dark spots of correct symmetry can also be assigned to the empty sites of Ta once Ta atoms are ordered as shown in Figure 2(d).

The first-principle calculations were carried out using the QUANTUM ESPRESSO code.[15] The energy cutoff for the plane wave expansion was 40 Ry and the Brillouin zone (BZ) was sampled by a $12 \times 12 \times 6$ and $12 \times 12 \times 4$ mesh for the low (*P$\bar{6}$m2*) and high (*P6/mmm*) symmetry structures. Spin-orbit (SO) coupling was included in all calculations. The band structures of PbTaSe$_2$ with low and high symmetries are shown in Figure 5. As shown in Figure. 5(a), the low-symmetry phase ($P\bar{6}m2$) lacks the inversion symmetry for Ta atoms. In order to preserve the inversion symmetry, the Ta atoms must be arranged to occupy two different sites along the *c*-axis alternatively. Hence, the unit cell used in the calculation doubles in the *c*-direction. Our results for the low-symmetry ($P\bar{6}m2$) structure (Fig. 5(a)) are in good agreement with those published in the literature. [16] Large SO splitting is seen at K and H points for both the Pb and Ta orbitals. We recently predicted that the electron-like Pb-6p orbital and hole-like Ta-5d orbital cross each other at the Fermi level, and three spin full nodal-line states are formed at the K and H points (Fig. 5(c)). [17] It is remarkable to find that the spin-orbit nodal lines in PbTaSe$_2$ are not only protected by the reflection symmetry but also characterized by an integer topological invariant.[17] Figure 5(b) represents the band structure of the high-symmetry (*P6/mmm*) PbTaSe$_2$. At the firsrt glance, these two band structures are similar with only minor difference due to the different size of the BZ in the z-direction. However, zooming in around the K and H points, all nodal-line states are actually gapped for the high-symmetry (*P6/mmm*) (Fig. 5(c)-(d)) due to the missing reflection symmetry. The band crossing states can be gapped due to the SO coupling. It is clear that the growth of low symmetry ($P\bar{6}m2$) PbTaSe$_2$ is an important and critical step to study the nodal-line physics.

Figure 6 shows the field-cooled (FC) and zero-field-cooled (ZFC) magnetization curves for PbTaSe$_2$ single crystal measured in an applied magnetic field of 10 Oe, for both orientations of parallel (*H* ∥) and perpendicular (*H* ⊥) to the *ab*-plane. The magnetic susceptibility confirms the onset of a superconducting transition temperature of ~3.8 K, which is consistent to the recent report based on a polycrystalline sample.[16] Large diamagnetic screening is evidenced from the dimensionaless $\chi_{ZFC}$ value that approaches ~95% of the expected value of 1/4π for H perpendicular to the *ab*-plane before the geometric factor correction. The significant reduction of FC value below $T_c$ suggests the effect of flux pinning from defects, which is commonly observed in the superconductors with layered structure.[18] The observed strong anisotropy reflects the 2D transport network for a layered crystal structure with van der Waals gap between layers, similar to that observed in $NbSe_2$ of $T_c$ ~7.2 K with a similar layered structure.[19] The isothermal magnetization *M(H)* curves of $PbTaSe_2$ at *T* = 1.8 K for *H* ∥ and *H* ⊥ to the *ab* - plane are depicted in the inset of Figure 6. The hysteresis of the magnetization in the superconducting state resembles the typical behavior of a type-II superconductor. The lower critical field $H_{c1}$ is extrapolated from the deviation of the linear field dependence of *M(H)* in the Meissner state, i.e., ~65 Oe for *H* ⊥ and ~140 Oe for *H* ∥ to the *ab*-plane as shown in the inset of Figure 6. The upper critical fields $H_{c2}$ is determined from the field at the intersection between the superconducting and normal states in M(*H*). The $H_{c2}$ values for *H* ∥ and *H* ⊥ to the *ab*- plane are found to be ~1500 Oe and ~ 450 Oe for at *T* = 1.8 K, respectively.

We have measured electrical resistivity ρ(*T*) along different crystallographic axes ($ρ_x$ and $ρ_z$) of a single crystal sample in zero field and in applied magnetic field along *H* ⊥ to the *ab*- plane. Strong dependence on both *T* and *H* were found for the in-plane resistivity anisotropy, A∥ = $ρ_y$ /$ρ_x$, as well as for the out-of-plane resistivity anisotropy, A⊥ = $ρ_z$ /$ρ_x$. The in-plane $ρ_x$ (*I* ⊥ *H*) and out-of-plane $ρ_z$ (*I* ∥ *H*) electrical resistivity ρ(*T*) of single crystal $PbTaSe_2$ are shown in Figure 7(a). A metallic behavior is inferred from the nearly linear *T* dependence of ρ and the sharp resistance drop below ~10 K for both crystallographic directions. The inset of Figure. 7(a) shows the expanded view at low temperatures for both crystallographic directions. Sharp superconducting transition to the zero resistance state of transition width ~0.3 K with an onset temperature $T_C$~3.8 K is observed, which is in good agreement with that defined using the Meissner effect of susceptibility data shown in Figure 6.

The in-plane residual resistivity before entering the superconducting state is $\rho_{norm} \sim$ 0.28 µΩ-cm and with a residual resistivity ratio RRR = $\rho_{(300 K)}/\rho_{(5 K)} \sim$ 115, as shown in Figure 6(a). A small residual resistivity together with a large RRR value confirms the high quality of the as-grown single crystal sample, in contrast to the RRR = 6 obtained using the polycrystalline sample.[16] The normal state resistivity data are described by the Bloch-Gruneisen (BG) model [20] as

$$\rho(T) = \rho_0 + A\left(\frac{T}{\theta_R}\right)^5 \int_0^{\theta_R/T} \frac{x^5 dx}{(e^x-1)(1-e^{-x})} \tag{1}$$

The $\rho_x(T)$ (5 K ≤ T ≤ 300 K) data fit well with the BG model, as shown by the solid red curve in Figure 7(a), which implies that the dominant scattering mechanism is a simple phonon–electron scattering above ~ 5 K. The obtained fitting parameters are residual resistivity $\rho_0$ = 0.28 µΩ-cm, material dependent coefficient $A$ = 74.6 µΩ-cm, and a Debye temperature of $\theta_R$ = 96 K. The in-plane $\rho_x(T)$ data measured under various magnetic fields ($H$) are depicted in Figure 7(b). These data indicate that $T_c$ decreases with increasing $H$ and is suppressed to below ~2.2 K by 0.1 Tesla, which is consistent to the $H_{c2}$ = 0.15 Tesla estimated from the $M(H)$ at 1.8 K as shown in inset of Fig. 6.

The field dependence of $\rho_x$ is shown in Figure. 7(c) at various temperatures from 4 K to 2 K for $I \perp H$. We note that the magnetoresistance for field well above $H_{c2}$ follows a nearly $\sqrt{B}$ dependence (not shown). The upper critical field $H_{c2}$ is defined by the field at which $\rho$ = $\rho_{norm}$ ($H$ = 0)/2 for $I \perp H$ and $I \parallel H$, as shown in Figure 7(d). For $T$ < 3 K, $H_{c2}(T)$ is practically linear in low $T$ (blue dashed line) giving a slope of $dH_{c2}/dT$ = -0.77 Tesla/K for $I \perp H$, which is one order larger than that for $I \parallel H$. The $H_{c2}(T)$ data fitting using $H_{c2}(T) = H_{c2}(T = 0 K) [1-(T/T_c)^\alpha]^\alpha$ returns an exponent of $\alpha \approx 1.47$ and $H_{c2}(T=0 K)$ = 2.26 Tesla (red line) for $I \perp H$.[13, 21] The corresponding Ginzburg-Landau coherent length $\xi_{GL}$ ($T$ = 0 K) is estimated to be about 12 nm. The fitted exponent $\alpha$ is close to 1.5, which is in agreement with those extracted from several other non-centrosymmetric superconductors [16,22] inferred to be a non-BCS like pairing mechanism before. [21]

The heat capacity $C_p(T)$ of PbTaSe$_2$ single crystal sample measured at $H = 0$ and $H = 5$ Tesla are displayed in Figure 8(a), with the inset showing an expanded view for $T < 5$ K. The value of $C_p$ at 250 K is near 90 J/mol K, which is close to the Dulong-Petit high-T limit of the lattice heat capacity $C_V = 3nR = 99.8$ J/mole K within experimental error. The bulk nature of superconductivity in PbTaSe$_2$ is evidenced by a sharp jump of $C_p(T)$ curve near 2.8 K measured at $H = 0$. The critical temperature $T_c = 3.78$ K is taken as the midpoint of the $C_p(T)$ jump, which agrees fairly well with that deduced from both the magnetic susceptibility (Figure 6) and resistivity (Figure 7) measurements. The superconducting state is completely suppressed by external applied magnetic field of 5 Tesla, as expected when it is higher than the H$_{c2}$ (see the upper inset of Figure 8(a)). The $C_P(T)$ data is fitted using $C_P(T)=\gamma T+\beta T^3$ and the plot of $C_p(T)/T$ versus $T^2$ linear fit (lower inset of Figure. 8(a)) yields the electronic specific heat coefficient (Sommerfeld coefficient) of $\gamma = 5.96$ mJ mol$^{-1}$ K$^{-2}$ and phonon specific heat coefficient of $\beta = 2.69$ mJ mol$^{-1}$ K$^{-4}$. Debye temperature $\Theta_D = 144$ K can be derived from $\beta$ using the relationship of $\Theta_D = (12\pi^4 Rn/5\beta)^{1/3}$, where $R$ is the molar gas constant and $n = 4$ is the number of atoms per formula unit. The value of Debye temperature $\Theta_D$ is higher than that of $\Theta_R = 96$ K estimated from the $\rho(T)$ data under the BG model, which has been suggested coming from the different assumptions of the Debye and BG models.[23]

We find the derived ratio of $\Delta C_p/\gamma T_c = 1.46$ is close to the 1.43 given by the BCS theory for weakly coupled superconductors.[8] In the superconducting state, electronic specific heat ($C_e(T)=C_p(T)-\beta T^3$) follows a single gapped BCS like function $C_e(T) = A\gamma T_c \exp(-\Delta_0/k_B T)$. Figure 8(b) shows the plot of $lnC_e(T)$ versus $1/T$ with the fitted exponential function in solid line. The fitting parameters are A = 17.6 and energy gap $\Delta_0/k_B$ = 6.69 K. The obtained energy gap is comparable to the limit of $1.76k_BT_c$ of weakly coupled BCS superconductors.[8] In order to get some information on the strength of the electron-phonon coupling, we estimate the average electron-phonon coupling constant $\lambda_{el-ph}$ using McMillan's theory [24]

$$\lambda_{el-ph} = \frac{1.04 + \mu^* \ln(\Theta_D/1.45\,T_c)}{(1 - 0.62\mu^*)\ln(\Theta_D/1.45\,T_c) - 1.04} ,$$

where $\mu^*$ is the Coulomb repulsive screened parameter, usually assumed to be between 0.1 and 0.15. [24,25] Substituting $T_c$ = 3.73 K, $\Theta_D$ = 144 K, and $\mu^*$ = 0.13 into above equation yields $\lambda_{el-ph}$ = 0.74, which is suggestive of a moderate electron-phonon coupling in PbTaSe$_2$. All of the extracted parameters shown above, including the superconducting anisotropy, the value of the specific heat jump $\Delta C_p/\gamma T_c$ at $T_c$, the superconducting energy gap $\Delta_0/k_B$, and the electron-phonon coupling $\lambda_{ep}$ support that PbTaSe$_2$ can be categorized as a weakly coupled type-II superconductor.

### 3. Conclusion

A high-quality single crystal of hexagonal PbTaSe$_2$ (space group $P\bar{6}m2$) was synthesized for the first time using a chemical vapor trasport technique. Although XRD cannot distinguish the space group has centrosymmetry or not conclusively, STM surface profile scan suggests that the highly ordered atomic plane is consistent to a crystal symmetry of noncentrosymmetric spacegroup ($P\bar{6}m2$). The bulk superconductivity below 3.83±0.02 K is confirmed with electrical resistivity $\rho(T)$, magnetic susceptibility $\chi(T)$, and specific heat measurements. The parameters of electron-phonon coupling $\lambda_{ep}$, energy gap $\Delta_0/k_B$, and the specific heat jump $\Delta C/\gamma T_c$ at $T_c$ imply that PbTaSe$_2$ can be ascribed to a weakly coupled type-II superconductor. Electronic band structure calculations show a complex Fermi surface and a moderately high DOS at the Fermi level.

## 4. Experimental Section

Chemical vapor transport method was employed with $PbCl_2$ as the transport agent for the $PbTaSe_2$ crystal growth, which allows an effective and faster vapor transport to produce the necessary super saturation of the expected final product. A three-zone muffle furnace was fabricated for this purpose, having a typical temperature profile as shown in Figure 9. Stoichiometric amount of Pb:Ta:Se (6N purity for Pb and Ta and 5N for Se) was loaded into a quartz ampoule, evacuated, sealed and fed into a furnace (850°C) for 5 days. About 10g of the pre-reacted $PbTaSe_2$ powder was placed together with a variable amount of $PbCl_2$ (purity 5N) (50 to 75mg) at one end of the silica ampoule (length 30-35 cm with inner diameter of 2.0 cm and outer diameter of 2.5 mm). All treatments were carried out in a glove box, with continuous purification of the Argon gas of oxygen and water level kept below 1 ppm. The loaded ampoule was evacuated, sealed and fed into a furnace for the growth next. The end of the ampoule containing the pre reacted material was held at 850°C, while the crystals grew at the other end of the ampoule was maintained at a temperature of 800°C (i.e., a temperature gradient near 2.5 K/cm) for about one week. Shiny single crystals of sizes up to $7 \times 5 \times 2$ mm$^3$ were obtained. The growth habit of the crystals was characterized by the dominance of (00*l*) faces, as confirmed by the Laue picture and XRD with preferred, as shown in the inset of Figure 1(a). Chlorine concentrations in the range of 2.6-3.7 mg $PbCl_2$/cm$^3$ yielded sufficiently high transport rates of 40-50 mg per day. Debye-Scherrer measurements of the lattice constants *a* and *c* at room temperature for samples taken from the pre-reacted polycrystalline $PbTaSe_2$ (a = 3.448(3) Å, c = 9.352(1) Å) as well as from the grown single crystals (a = 3.4474(2) Å, c = 9.3528(3) Å) showed good agreement with the literature values. [14]

**Acknowledgements**

RS and FCC acknowledge the support provided by the Academia Sinica research program on Nanoscience and Nanotechnology under project number NM004. FCC acknowledges support from Ministry of Science and Technology in Taiwan under project number MOST-102-2119-M-002-004.

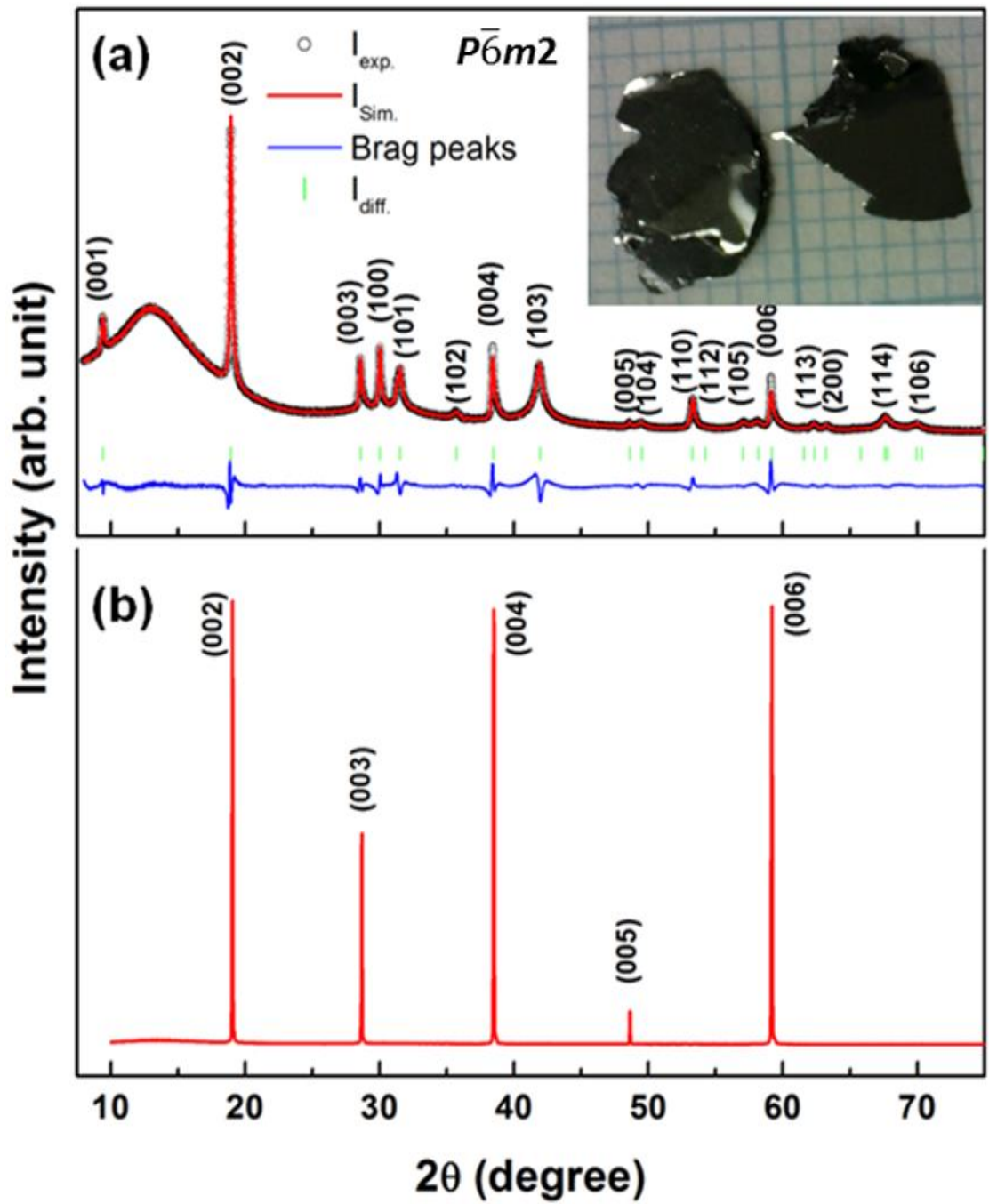

Fig.1(a). The powder x-ray diffraction (XRD) pattern of the as-grown sample. In the inset shows a photograph of PbTaSe$_2$ single crystals grown with the chemical vapor transport method. (b) Single crystal diffraction pattern obtained using an x-ray beam to show the preferred orientation of (00$l$) plane.

**Higher symmetry (*P6/mmm*)**

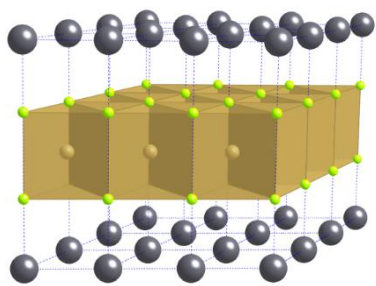
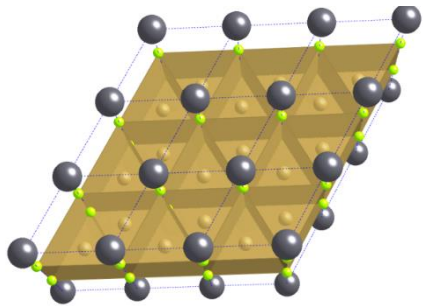

(a)                (b)

**Lower symmetry (*P6̄m2*)**

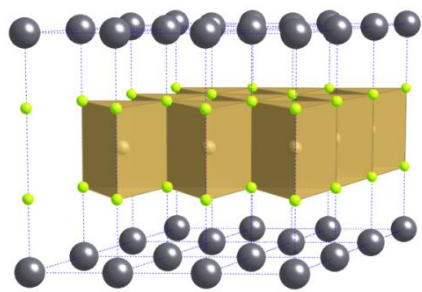
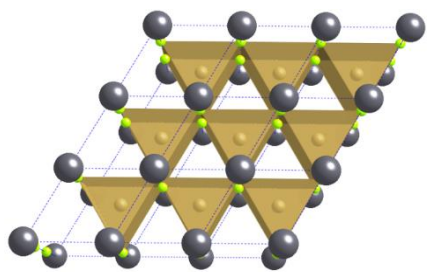

(c)                (d)

Fig 2. Crystal structures of PbTaSe$_2$ with *P6/mmm* (a,b) and *P6̄m2* (c,d) symmetries.

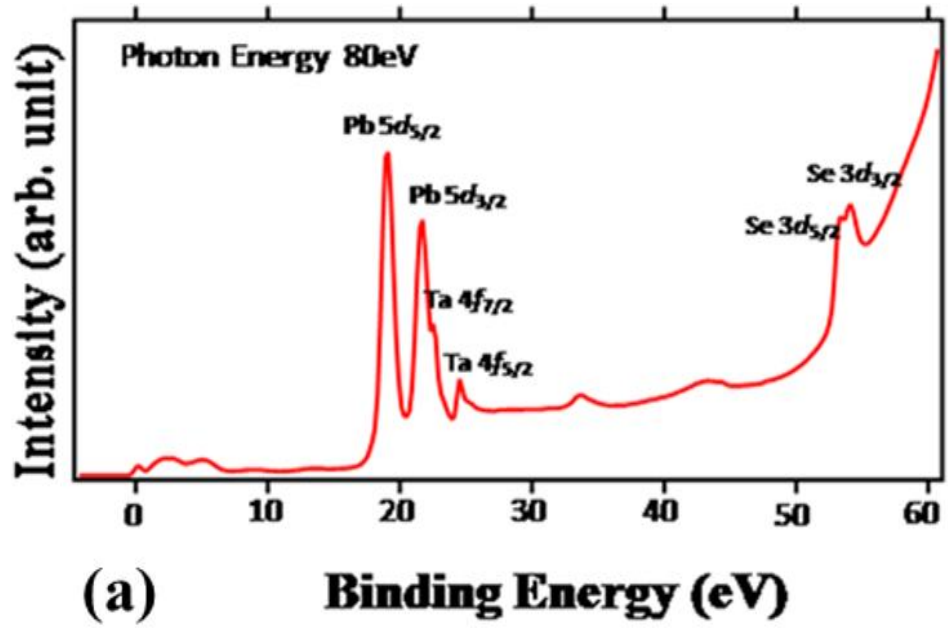

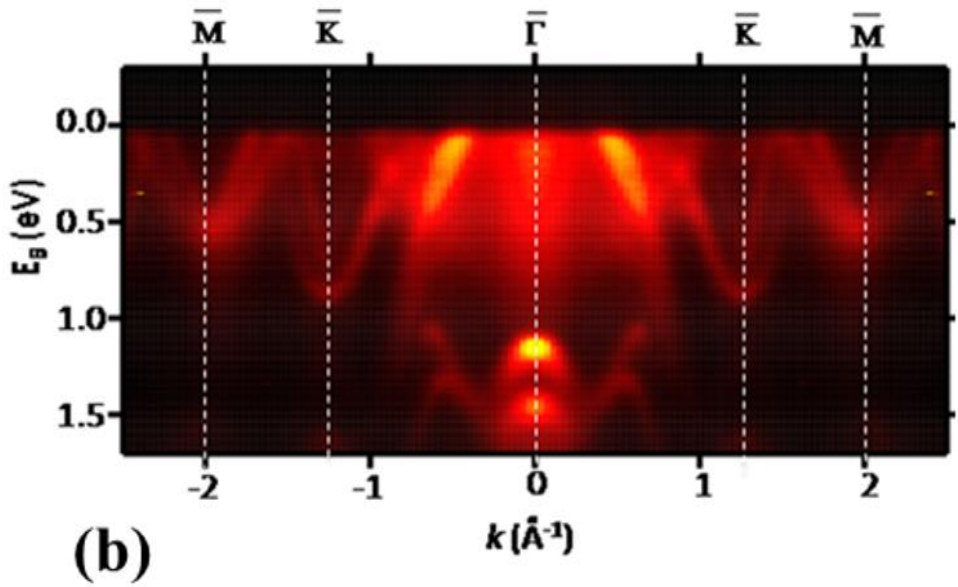

Fig. 3. (a) ARPES core level spectrum showing clear Pb 5d, Se 3d and Ta 4f core level peaks. (b) The ARPES spectra taken along $\bar{M}-\bar{K}-\bar{\Gamma}$ with 64 eV photons near the Fermi level.

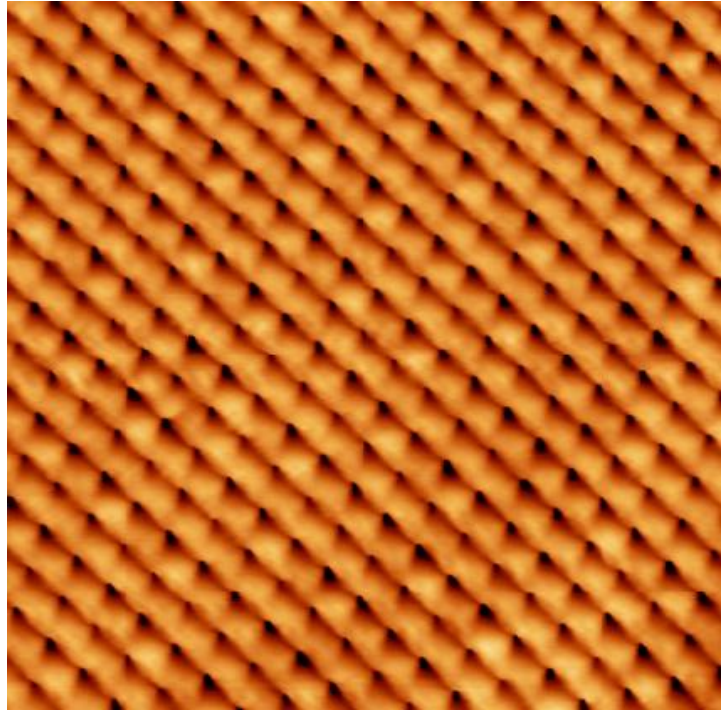

Fig. 4. The STM image (4.8x4.8nm$^2$) obtained with bias voltage -0.5 V and tunneling current 0.7 nA.

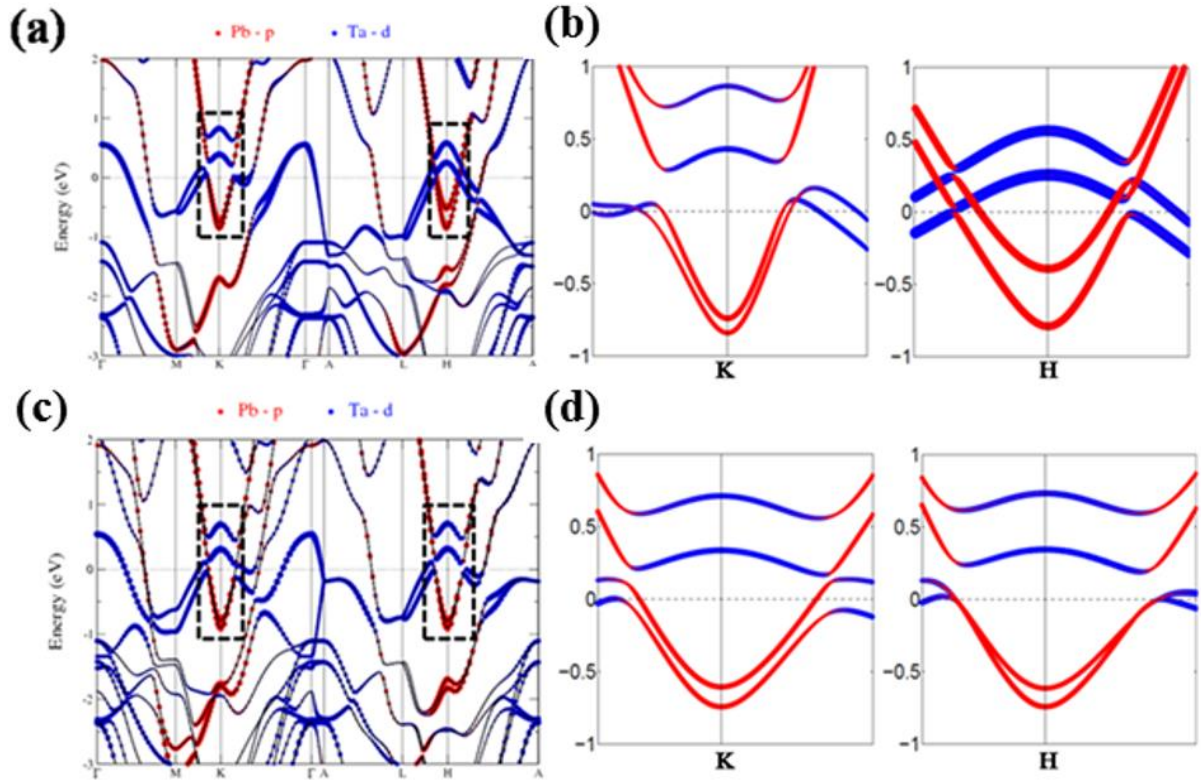

Fig. 5. Band structures of PbTaSe$_2$ calculated with (a) non-centrosymmetric ($P\bar{6}m2$) and (c) centrosymmetric (*P6/mmm*) space groups. (b) and (d) are the zoom-in picture near the K and H points for the low symmetry ($P\bar{6}m2$) and high symmetry (*P6/mmm*) structures, respectively.

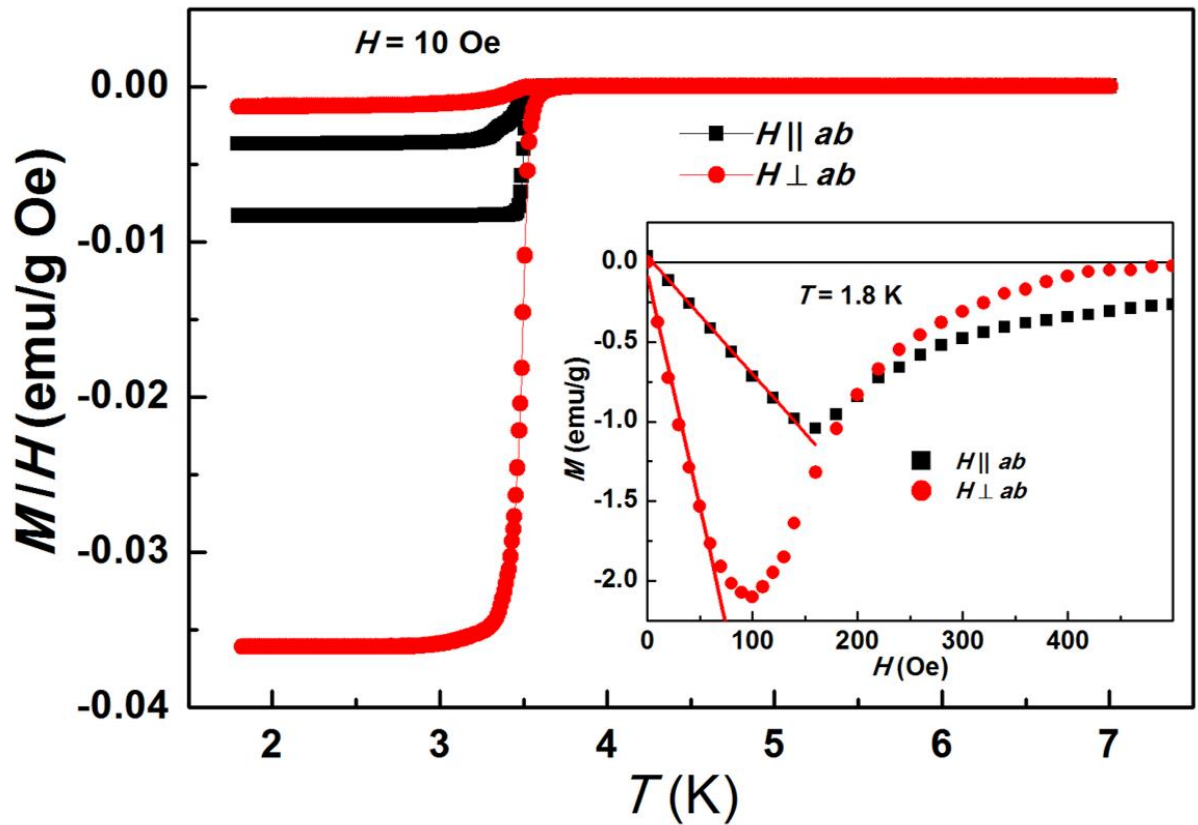

Fig. 6. Zero-field cooled and field cooled dc-magnetization at $H$ = 10 Oe measured for both $H \parallel$ and $H \perp$ to the *ab*-plane as a function of temperature. Inset shows the isothermal magnetization measured at 1.8 K for both $H \parallel$ and $H \perp$. The solid lines are linear fits to the data to extract $H_{c1}$.

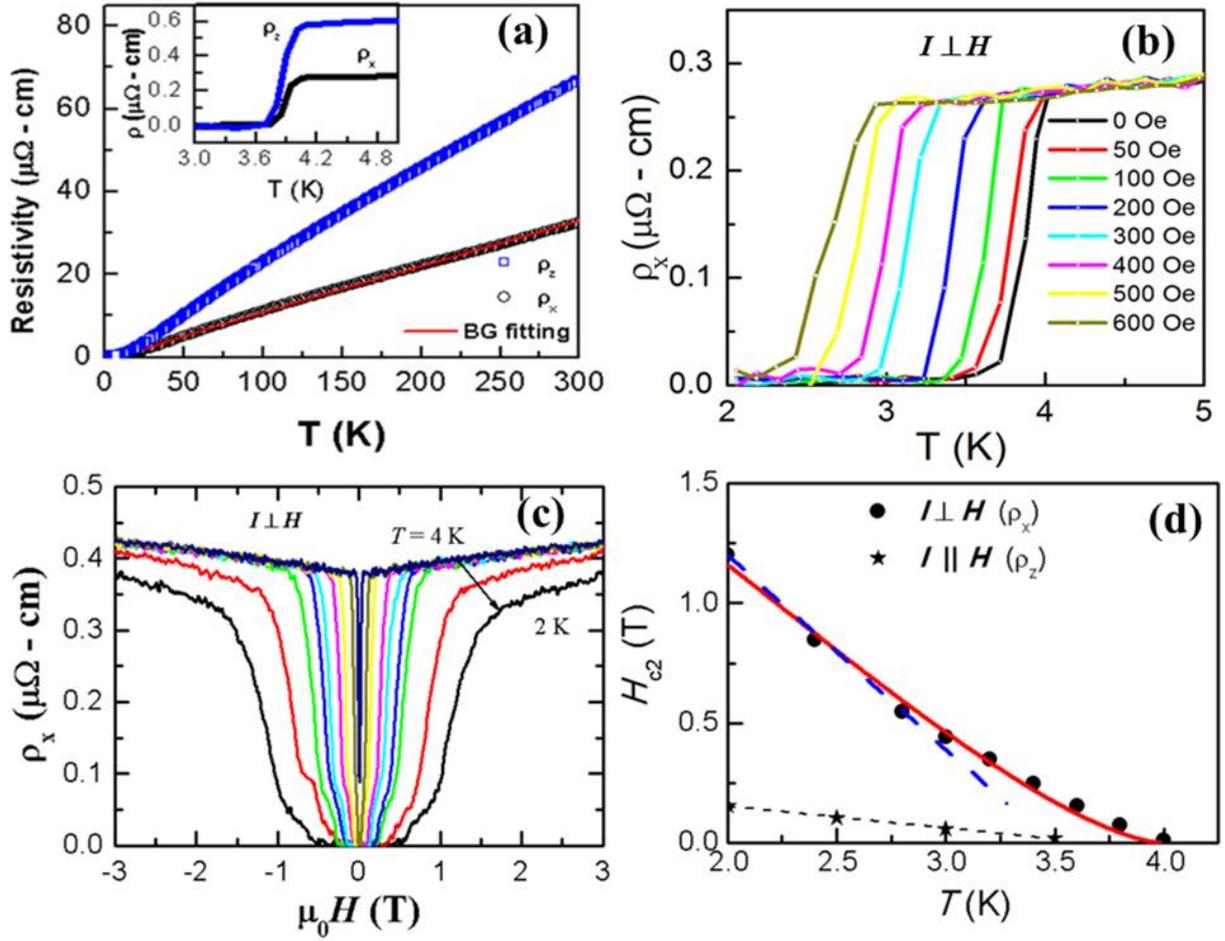

Fig.7 (a) ρ–T curves of a single crystal PbTaSe$_2$ for both in-plane (ρ$_x$) and out-of-plane (ρ$_z$). The inset figure shows the superconducting transition with vanishing ρ below T = 3.8 K for both crystallographic directions. (b) ρ$_x$–T curves at various H below 5 K for I ⊥ H. (c) ρ$_x$ –H curves at various T ranging from 2 K to 4 K for I ⊥ H. The T dependence of the upper critical field H$_{c2}$ for both I ⊥ H and I ∥ H are plotted in (d). A fitting function of $H_{c2}(T) = H_{c2}(0) [1-(T/T_c)^\alpha]^\alpha$ is used (red line) with fitting parameters of exponent α ≈ 1.47 and H$_{c2}$(0) = 2.26 Tesla. The blue dashed line is a linear fit to the low T data of a slope dH$_{c2}$/dT = -0.77 T/K.

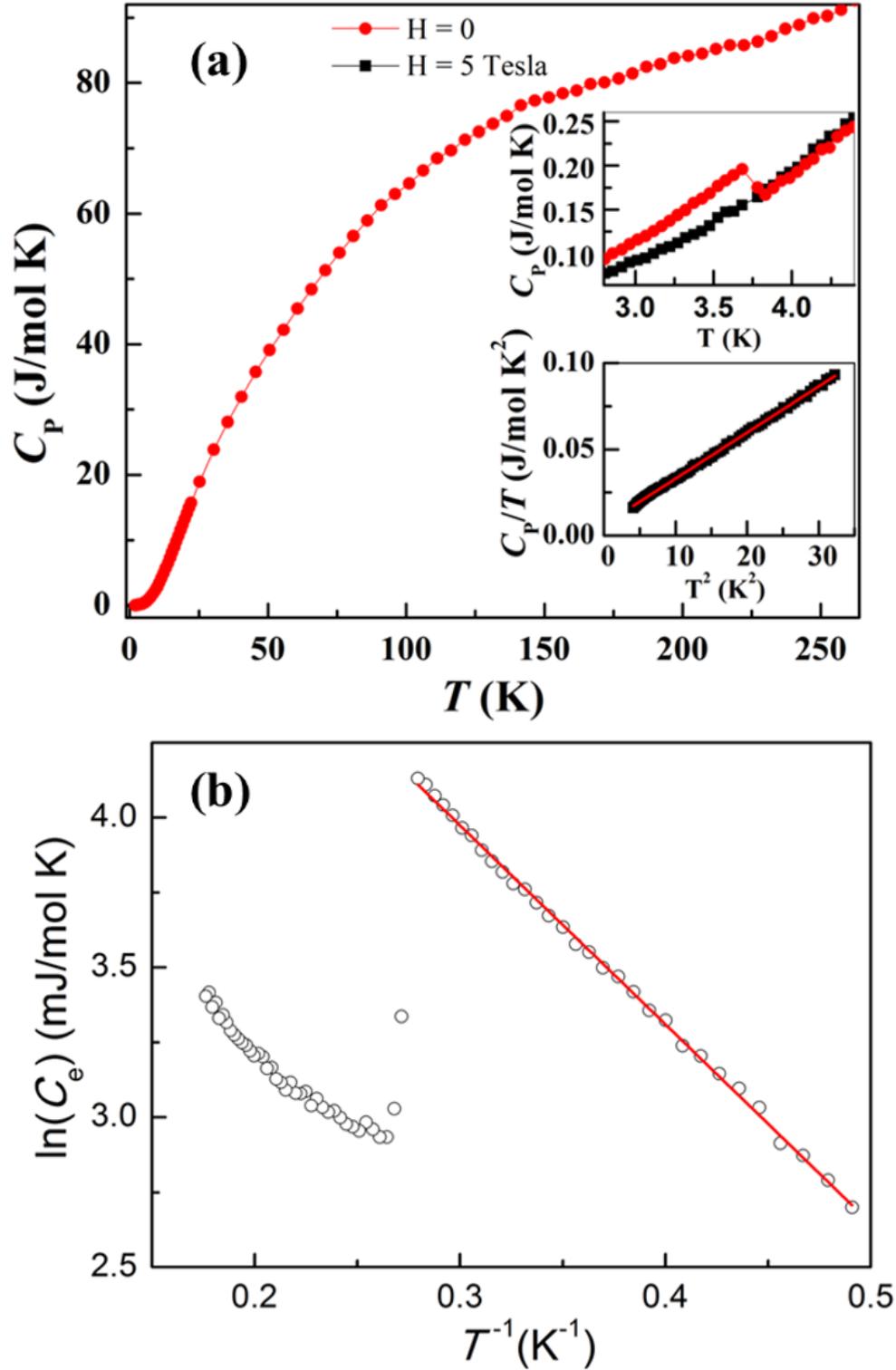

Fig.8(a). Heat capacity $C_p$ of PbTaSe$_2$ as a function of temperature $T$ measured in zero magnetic field (main panel). The upper inset shows the low temperature regime under H = 0 and 5 Tesla. The lower inset shows $C_p/T$ versus $T^2$ in H = 5 T. The solid red line presents the linear fitting as described in the text. (b) The electronic specific heat in logarithmic scale is plotted as a function of 1/T. The solid red line indicates the exponential function given by the BCS theory, see text for details.

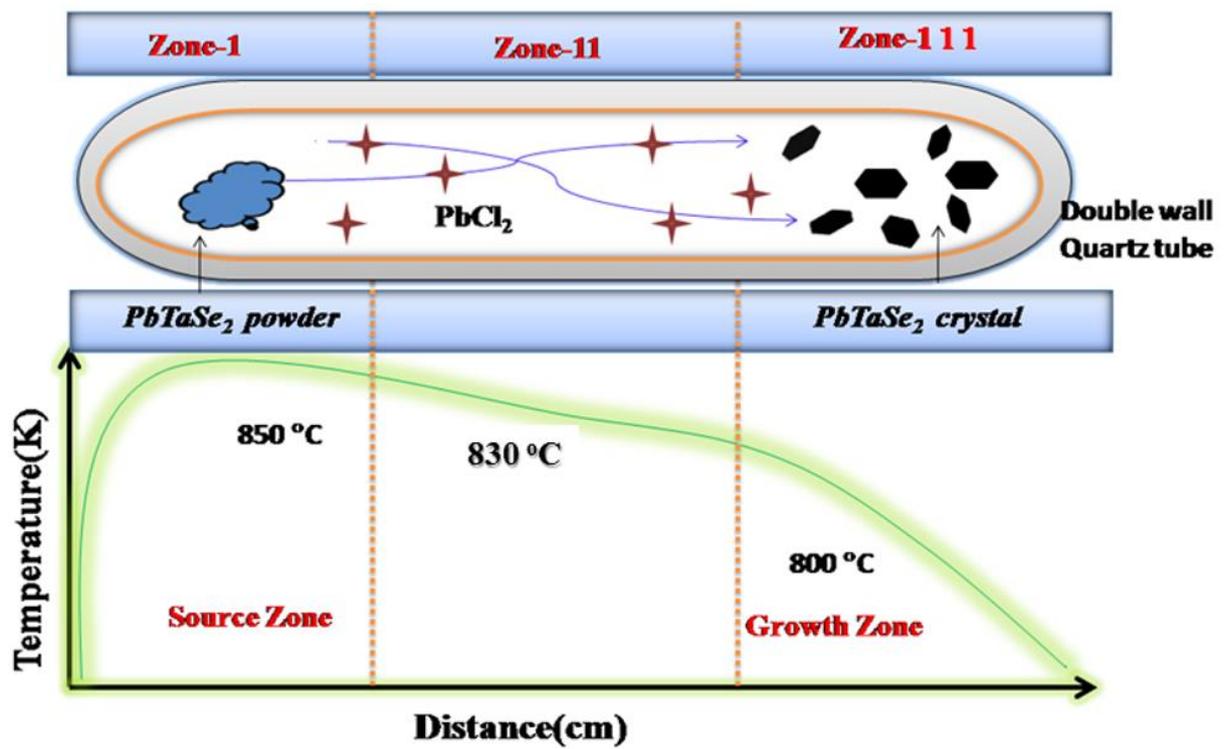

Fig. 9 Cross-sectional side view and temperature profile of a double wall quartz tube sitting in a 3-zone tube furnace used for chemical vapor transport growth of PbTaSe$_2$ single crystal. The zone-I and zone III were set at 850 and 800 C, respectively.

**Table 1.** List of refined structural parameters of the PbTaSe$_2$ sample for both space groups, where B$_{iso}$ represents the isotropic temperature parameter.

**PbTaSe$_2$ (Hexagonal)**

$P\bar{6}m2$ space group, $a = b = 3.447$ Å and $c = 9.3527$ Å

| Atom | x | y | z | B$_{iso}$ (Å$^2$) | Occupancy |
|---|---|---|---|---|---|
| Pb | 0 | 0 | 0 | 2.41 | 1.00 |
| Ta | 0.3333 | 0.6667 | 0.5 | 1.78 | 1.00 |
| Se | 0 | 0 | 0.31 | 2.29 | 1.00 |

$P6/mmm$ space group, $a = b = 3.4474$ Å and $c = 9.3528$ Å

| Atom | x | y | z | B$_{iso}$ (Å$^2$) | Occupancy |
|---|---|---|---|---|---|
| Pb | 0 | 0 | 0 | 1.24 | 1.00 |
| Ta | 0.3333 | 0.6667 | 0.5 | 2.35 | 0.50 |
| Se | 0 | 0 | 0.31 | 1.42 | 1.00 |